\documentclass[pra,twocolumn,floatfix]{revtex4}
\usepackage{graphicx}

\newlength{\onecolfig}
\newlength{\twocolfig}
\newcommand{\ion}[2]{\mbox{$^{#2}$#1$^+$}}
\newcommand{\Ca}[1]{\ion{Ca}{#1}}

\newcommand{\Sr}[1]{\ion{Sr}{#1}}
\newcommand{\Be}[1]{\ion{Be}{#1}}
\newcommand{\Mg}[1]{\ion{Mg}{#1}}
\newcommand{\lev}[2]{\mbox{#1$_{\mbox{\tiny$#2$}}$}}
\newcommand{\fslev}[3]{\mbox{#1$^{\mbox{\tiny$#3$}}_{\mbox{\tiny$#2$}}$}}
\newcommand{\hfslev}[3]{\mbox{#1$^{\mbox{\tiny$#3$}}_{\mbox{\tiny$#2$}}$}}

\newcommand{\unit}[1]{\,\mbox{#1}}

\newcommand{\kHz}{\unit{kHz}}
\newcommand{\MHz}{\unit{MHz}}
\newcommand{\GHz}{\unit{GHz}}
\newcommand{\THz}{\unit{THz}}

\newcommand{\mW}{\unit{mW}}

\newcommand{\um}{\unit{$\mu$m}}
\newcommand{\nm}{\unit{nm}}

\newcommand{\us}{\unit{$\mu$s}}

\newcommand{\mT}{\unit{mT}}
\newcommand{\uT}{\unit{$\mu$T}}

\newcommand{\etal}{{\em et al.}}

\newcommand{\ish}{\mbox{$\sim$}\,}

\newcommand{\ket}[1]{\mbox{$\left| #1 \right>$}}

\newcommand{\sub}[1]{\mbox{$_{\mbox{\tiny #1}}$}}

\begin{document}

\title{Hybrid quantum logic and a test of Bell's inequality\\ using two different atomic isotopes}

\author{C. J. Ballance, V. M. Sch\"afer, J. P. Home, D. J. Szwer, S. C. Webster, D. T. C. Allcock,\\ N. M. Linke, T. P. Harty, D. P. L. Aude Craik, D. N. Stacey, A. M. Steane and D. M. Lucas}

\affiliation{Department of Physics, University of Oxford, Clarendon Laboratory, Parks Road, Oxford OX1 3PU, U.K.}

\date{27 November 2015, v40.43}

\maketitle

{\bf Entanglement is one of the most fundamental properties of quantum mechanics~\cite{Schrodinger1935,Einstein1935,Bell1964}, and is the key resource for quantum information processing~\cite{Deutsch1985,Ekert1991}. Bipartite entangled states of identical particles have been generated and studied in several experiments, and post-selected or heralded entangled states involving pairs of photons, single photons and single atoms, or different nuclei in the solid state, have also been produced~\cite{Freedman1972,Aspect1982,Rowe2001,Moehring2004,Giustina2013,Christensen2013,Pfaff2013}. Here, we use a deterministic quantum logic gate to generate a ``hybrid'' entangled state of two trapped-ion qubits held in different isotopes of calcium, perform full tomography of the state produced, and make a test of Bell's inequality with non-identical atoms. We use a laser-driven two-qubit gate~\cite{Leibfried2003}, whose mechanism is insensitive to the qubits' energy splittings, to produce a maximally-entangled state of one \Ca{40} qubit and one \Ca{43} qubit, held 3.5\um\ apart in the same ion trap, with 99.8(6)\% fidelity. We test the Clauser-Horne-Shimony-Holt~\cite{Clauser1969} (CHSH) version of Bell's inequality for this novel entangled state and find that it is violated by 15 standard deviations; in this test, we close the detection loophole~\cite{Rowe2001} but not the locality loophole~\cite{Aspect1982}. Mixed-species quantum logic is a powerful technique for the construction of a quantum computer based on trapped ions, as it allows protection of memory qubits while other qubits undergo logic operations, or are used as photonic interfaces to other processing units~\cite{Wineland1998a,Monroe2013}. The entangling gate mechanism used here can also be applied to qubits stored in different atomic elements; this would allow both memory and logic gate errors due to photon scattering to be reduced below the levels required for fault-tolerant quantum error correction, which is an essential pre-requisite for general-purpose quantum computing.
}

For Schr{\"o}dinger, entanglement was ``{\em the} characteristic trait of quantum mechanics''~\cite{Schrodinger1935} and it has been at the heart of debates about the foundations of quantum mechanics since the framing of the Einstein-Podolsky-Rosen paradox~\cite{Einstein1935}. The theoretical work of Bell~\cite{Bell1964}, and of Clauser \etal~\cite{Clauser1969}, established an experimental test which could be used to rule out local hidden-variable theories on the basis of correlations between measured properties of entangled particles, and numerous experiments, starting with that of Freedman and Clauser, have confirmed the predictions of quantum mechanics~\cite{Freedman1972,Aspect1982,Rowe2001,Moehring2004,Giustina2013}. Tests of Bell's inequality with trapped ions were the first to close the so-called ``detection loophole''; hitherto these trapped-ion tests have exclusively been carried out with identical atoms~\cite{Rowe2001,Matsukevich2008,Lanyon2014}. The entanglement explored in tests of Bell's inequality is typically an entanglement between distinguishable particles, in the strict quantum mechanical sense, but when the particles are identical in their internal structure and state, they are distinguishable only through their spatial localization. By employing different isotopes, our experiments involve entities that are also distinguishable by many internal properties, such as baryon number, mass, spin, resonant frequencies, and so on.

Apart from its intrinsic interest, entanglement is a central resource for quantum information applications, such as quantum cryptography~\cite{Ekert1991} and quantum computing~\cite{Deutsch1985}. Trapped atomic ions are one of the most promising technologies for the implementation of quantum computation; several demonstrations of simple multi-qubit algorithms have been made~\cite{Blatt2008} and the elementary set of quantum logic operations has recently been demonstrated with the precision required for the implementation of fault-tolerant techniques~\cite{Harty2014,Ballance2014b}. Scaling up trapped-ion systems to the large numbers of qubits required for useful quantum information processing and quantum simulation will almost certainly require the use of more than one species of ion, both for the purpose of sympathetic laser-cooling (which allows independent control of the external and internal atomic degrees of freedom)~\cite{Wineland1998a,Barrett2003,Home2009a} and for providing robust memory qubits. The best memory qubits reside in hyperfine ground states~\cite{Langer2005,Harty2014}, which have essentially infinite lifetimes against spontaneous decay, but are vulnerable to the scattering of a single photon of resonant laser light. In a complex, multi-zone, ion trap processor it will be difficult to shield the memory qubits sufficiently well from resonant laser beams, hence it will be useful to employ different species of ion, for example as memory and logic qubits, and a high-fidelity entangling gate operation between the two species will be invaluable. A significant initial step was the demonstration of coherent state transfer between different species in the context of precision metrology~\cite{Schmidt2005,Hume2007}. The relative merits of using different isotopes versus different elements are discussed below. 

\begin{figure}
\includegraphics[width=\onecolfig]{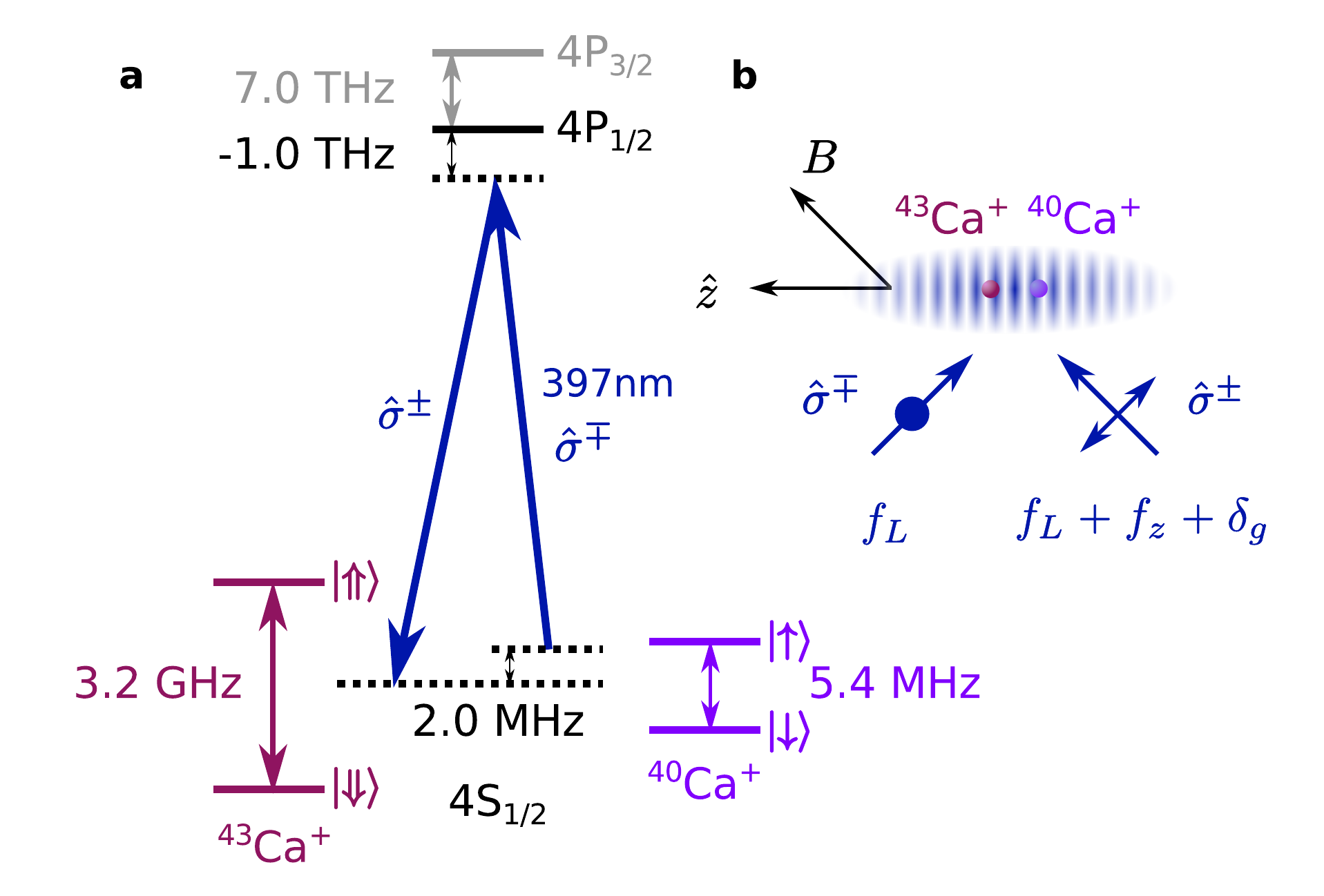}
\caption{%
Calcium ion energy levels and experimental geometry. 
(a) Qubit states and Raman transitions in \Ca{43} and \Ca{40}. The two Raman beams have a mean detuning of $\Delta=-1.04\THz$ from the $\lev{4S}{1/2}\leftrightarrow\lev{4P}{1/2}$ ($397\nm$) transition, and a difference frequency of $\delta=f_z+\delta_g \approx 2\MHz$.
(b) Raman gate beam geometry. The two perpendicular beams are aligned to set the lattice $k$-vector parallel to the trap axis $\hat{z}$. The beams have waist radii $w=27\um$, a power of $\approx 5\mW$ each, and orthogonal linear polarizations as indicated. A third, $\pi$-polarized, Raman beam (not shown) co-propagates with the $\sigma^\mp$ beam and is used for sub-Doppler sideband cooling and single-qubit operations on \Ca{40}. The quantization axis is set by a magnetic field $B\approx 0.2\mT$. The diagram is not to scale: the ions are separated by 3.5\um, which is $12\frac{1}{2}$ periods of the standing wave, and around 20,000 times the atomic radius of calcium.
}
\label{F:levels}
\end{figure}

In the present work, we entangle qubits stored in two different isotopes of calcium. The \Ca{40} qubit is stored in the Zeeman-split ground level, $(\ket{\downarrow},\ket{\uparrow})=(\fslev{4S}{1/2}{-1/2},\fslev{4S}{1/2}{+1/2})$, and the \Ca{43} qubit is stored in the hyperfine ground states $(\ket{\Downarrow},\ket{\Uparrow})=(\hfslev{4S}{1/2}{4,+4},\hfslev{4S}{1/2}{3,+3})$, see figure~\ref{F:levels}. The qubit energy splittings differ by some three orders of magnitude ($f_\updownarrow\approx 5.4\MHz, f_\Updownarrow\approx 3.2\GHz$), but they may nevertheless be efficiently coupled via the two-qubit gate mechanism of Leibfried~\etal~\cite{Leibfried2003}, in which the ``travelling standing wave'' from a pair of far-detuned laser beams exerts a qubit-state-dependent force on the ions whose magnitude $F$ is largely independent of the qubit frequency. The force originates from a spatially-varying light shift, oscillates at the difference frequency $\delta$ between the two beams and, when $\delta=f_z+\delta_g$ is set close to the resonant frequency $f_z$ of a normal mode of motion of the two-ion crystal, a two-qubit phase gate may be implemented by applying the force for a time $(1/\delta_g)$. An advantage of this type of gate is that the phase of the optical field does not need to be referenced to either of the qubit phases (see Methods); this makes scaling the system easier because the relative optical phase does not need to be controlled between different trap zones, or during time delays between gates.

\begin{figure}
\includegraphics[width=\onecolfig]{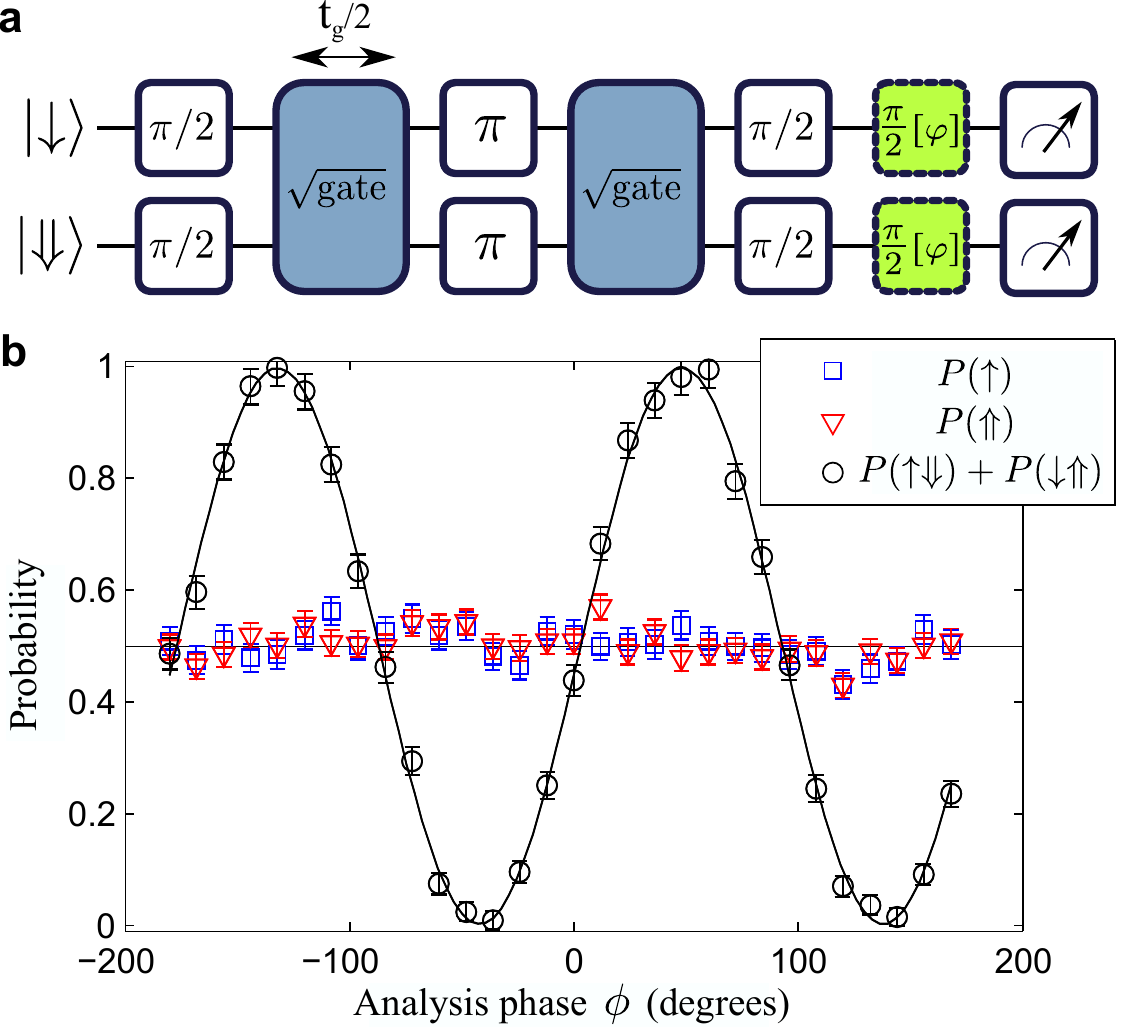}
\caption{%
Entangling gate sequence and results.
(a) Gate sequence, showing the operations applied to the \Ca{40} (upper line) and \Ca{43} (lower line) qubits. The final state analysis (tomography) $\pi/2$-pulses shown in green are optional; by scanning their phase $\phi$ we can diagnose the state produced by the gate. 
(b) Qubit populations and parity signal after correcting for readout errors (see Methods). The individual qubit populations are consistent with $\frac{1}{2}$, as expected for the Bell state $(\ket{\downarrow\Downarrow}+\ket{\uparrow\Uparrow})/\sqrt{2}$. The parity signal $P(\uparrow\Downarrow)+P(\downarrow\Uparrow)$, i.e.\ the probability of the two qubits being in opposite states, should oscillate between 0 and 1 as $\sin(2\phi)$ for a perfect Bell state. From the contrast of the parity signal and a measurement of the populations without the analysis pulses, we infer a Bell state fidelity of 99.8(6)\%. The error bars show $1\sigma$ statistical errors. 
}
\label{F:parity}
\end{figure}

An important difference in the gate mechanism compared with the case of identical ions is that the forces on corresponding qubit states differ ($F_\uparrow\neq F_\Uparrow$ and $F_\downarrow\neq F_\Downarrow$) so that, in general, the four possible qubit states ($\uparrow\Uparrow$, $\uparrow\Downarrow$, $\downarrow\Uparrow$, $\downarrow\Downarrow$) each acquire different phases. We choose to implement the gate operation in two halves, each of duration $t_g/2=1/\delta_g$, separated by spin-flip operations ($\pi$-pulses) on the qubits (figure~\ref{F:parity}a). This symmetrizes the gate operation such that the relative phases acquired by the four states are $(0,\Phi,\Phi,0)$. By setting the laser power (i.e., effective Rabi frequency) and gate detuning $\delta_g$ appropriately, such that $\Phi=\pi/2$, and enclosing the gate operation in a Ramsey interferometer (two pairs of $\pi/2$-pulses), we can generate the maximally-entangled Bell state $(\ket{\downarrow\Downarrow}+\ket{\uparrow\Uparrow})/\sqrt{2}$ from the initial state $\ket{\downarrow\Downarrow}$. The $\pi$-pulses also protect the qubits against dephasing due to slow ($\gg t_g$) variations in magnetic fields.

In our experiment, we implement the gate using the in-phase axial motional mode (at $f_z=2.00\MHz$) of a linear Paul trap~\cite{Ballance2014a}, with the ion separation (3.5\um) equal to a half-integer number of standing wavelengths, thus exciting the motion maximally for the \ket{\uparrow\Downarrow} and \ket{\downarrow\Uparrow} states. The Lamb-Dicke parameters for the two different isotopes are $\eta_{40}=0.121$ and $\eta_{43}=0.126$. After initial Doppler cooling, both axial modes are cooled close to their ground states (mean occupation number $\bar{n}<0.1$) by Raman sideband cooling applied to the \Ca{40} ion, which sympathetically cools the \Ca{43} ion~\cite{Home2009}.  Both qubits are initialized by optical pumping, after which we apply the gate sequence shown in figure~\ref{F:parity}a, using a gate duration $t_g=27.4\us$. Single-qubit $\pi/2$- and $\pi$-pulses, for the spin-echo and tomography operations, are applied using co-propagating Raman beams (for \Ca{40}) and microwaves (for \Ca{43}). The ordering of the ion pair in the trap was kept constant over the time taken to acquire the full data set, to guard against systematic effects associated with ion position (see Methods). We implement individual single-shot qubit readout by state-selectively shelving both ions to the \lev{3D}{5/2} level simultaneously, then detecting the ions' fluorescence sequentially in two photomultiplier counting periods (see Methods).

\begin{figure}
\includegraphics[width=\onecolfig]{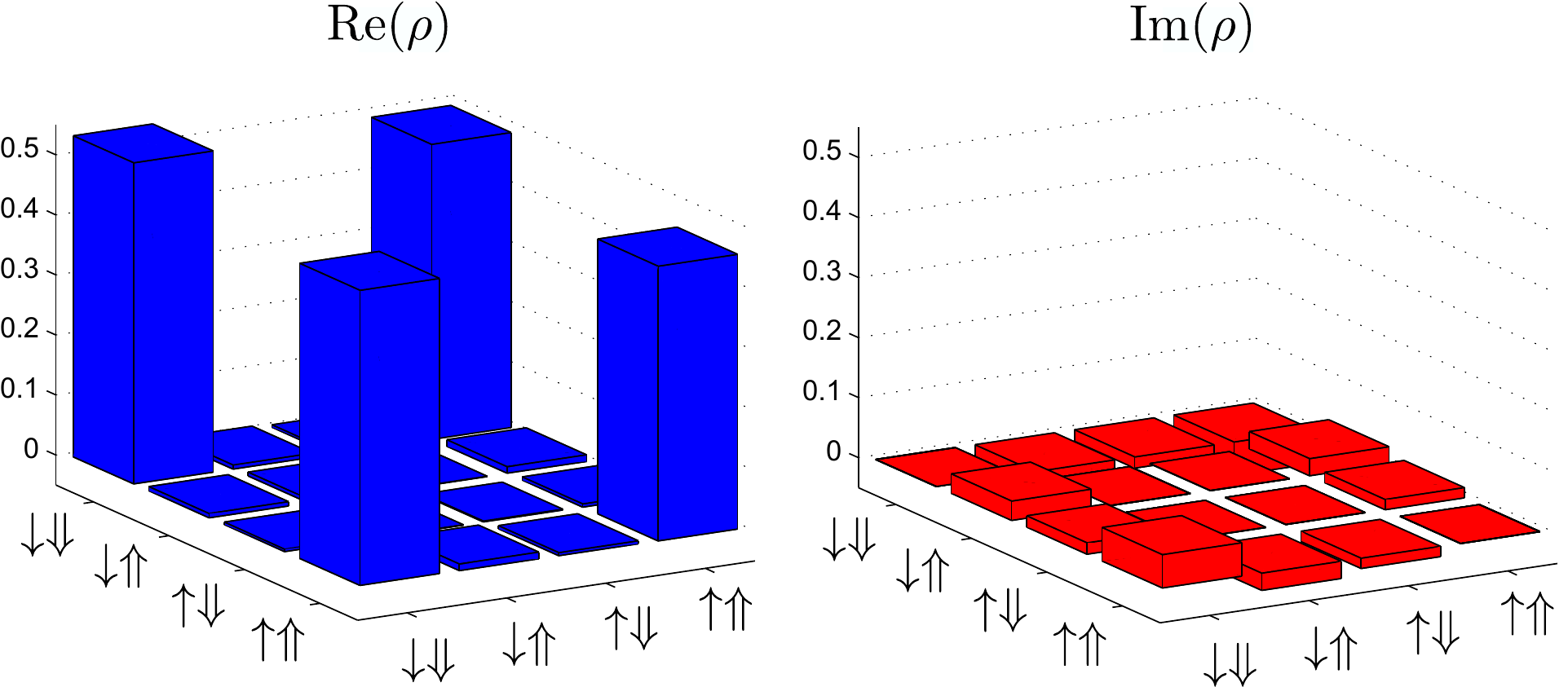}
\caption{%
Density matrix of the mixed-isotope Bell state. The plots show the real (left) and imaginary (right) parts of the density matrix, after correcting for qubit readout errors (see Methods). The measurements were made by rotating each qubit independently to perform full quantum state tomography. We used a maximum likelihood method to find the density matrix that best represents the experimental data. This gives a separate estimate of the gate fidelity, 99(1)\%.
}
\label{F:densitymatrix}
\end{figure}

From the contrast of the parity fringes shown in figure~\ref{F:parity}, and a measurement of the qubit populations before the analysis pulses~\cite{Leibfried2003}, we estimate the fidelity of the Bell state produced by the gate to be ${\cal F}=99.8(6)\%$, where the error is dominated by statistical uncertainty. Known contributions to the gate error are significantly smaller~\cite{Ballance2014a} than the statistical uncertainty; for example the photon scattering error at the $\Delta=-1.04\THz$ Raman detuning used is estimated to be $\approx 0.1\%$. Since the two qubits may be rotated independently by addressing them in frequency space, we can also perform full tomography of the entangled state and extract the density matrix (figure~\ref{F:densitymatrix}); the density matrix is consistent with that for the desired Bell state, to within the systematic errors from the imperfect tomography pulses, and gives a separate estimate of the fidelity ${\cal F}=99(1)\%$. In both cases, ${\cal F}$ represents the fidelity of the entangling gate operation; it excludes errors due to state preparation and readout, which we characterize in independent experiments (see Methods). 

To perform a test of the CHSH version of Bell's inequality, we follow the gate sequence with further independent single-qubit rotations and measurements. The single qubit rotations have constant phase $\phi$ but varying rotation angle $\theta$. From these measurements we determine the two-particle correlation functions with results shown in table~\ref{T:Bellparas}. As is well known, the maximal CHSH parameter $S$ allowed by local hidden-variable theories is 2, whereas quantum mechanics allows $S\leq 2\sqrt{2}$. In order to avoid having to make a fair-sampling assumption, we do not correct for qubit readout errors in these experiments. The finite detection error then limits the CHSH parameter to a detectable maximum $S\sub{max}=2.236(7)$ for a perfect Bell state; our results give $S=2.228(15)$, consistent with $S\sub{max}$ to within the stated uncertainties, and violating the CHSH inequality by $\approx 15\sigma$.

\begin{table}
\begin{tabular}{|c|cccc|}
 \hline
 $\theta_a$ (\Ca{40})   & $\pi/4$ & $3\pi/4$ & $\pi/4$ & $3\pi/4$ \\
 $\theta_b$ (\Ca{43})   & $\pi/2$ & $\pi/2$  & $0$     & $0$ \\
 \hline
 $E(\theta_a,\theta_b)$ & $0.565(7)$ & $0.530(7)$ & $0.560(7)$ & $-0.573(8)$ \\
 \hline
\end{tabular}
\caption{%
Bell/CHSH inequality test results, using the mixed-isotope entangled state. The qubits $a$ and $b$ are independently rotated through angles $(\theta_a,\theta'_a)=(\pi/4,3\pi/4)$ and $(\theta_b,\theta'_b)=(\pi/2,0)$, and for each combination of angles the correlation function $E(\theta_a,\theta_b)$ is measured, with results shown. ($E$ is defined as in ref.\cite{Matsukevich2008}.) The CHSH parameter is given by $S = |E(\theta_a,\theta_b)+E(\theta'_a,\theta_b)|+|E(\theta_a,\theta'_b)-E(\theta'_a,\theta'_b)| = 2.228(15) > 2$, thus violating Bell's inequality for this system of non-identical atoms. The state detection errors are sufficiently small ($\approx 6\%$, see Methods), that it is not necessary to make a fair-sampling assumption. For each angle setting 4,000 measurements were made.
}
\label{T:Bellparas}
\end{table}

The mixed-species quantum logic gate that we have demonstrated has allowed us to create a novel entangled state, leading to the first test of a Bell inequality violation between isolated non-identical atoms. As an application, the two isotopes used here could be employed for scalable quantum computing architectures based on trapped ions; hyperfine qubits in \Ca{43} at present constitute the best single-qubit memories ($T_2^*\,\ish 1\min$)~\cite{Harty2014}, whereas the simpler atomic structure of \Ca{40} is well suited for use as a ``photonic interconnect'' qubit~\cite{Monroe2013}. There are technical advantages to using ions of similar mass for sympathetic cooling and ion transport in multi-zone traps. However, while the relatively small isotope shifts ($\ish 1\GHz$) allow the convenient use of the same laser systems for manipulation of both species, they may provide insufficient protection of qubits from stray resonant light unless tightly focussed beams are used~\cite{Home2009,Lanyon2014}. Therefore in the long term it may be necessary to use different atomic elements~\cite{Barrett2003}. The gate mechanism employed here is independent of the qubit frequency and thus can also be used to couple qubits stored in different elements, provided that the Raman laser fields exert sufficient force on both qubits. We note that \Ca{} and \Sr{} ions are an attractive choice in this respect: the $\lev{4S}{1/2}\leftrightarrow\lev{4P}{1/2}$ transition in \Ca{} is separated from the $\lev{4S}{1/2}\leftrightarrow\lev{4P}{3/2}$ transition in \Sr{} by 20\THz. A Raman laser detuning $\Delta=-8\THz$ (comparable to that used in our recent \Ca{43}--\Ca{43} two-qubit gate experiments~\cite{Ballance2014b}) would enable the implementation of a mixed-species logic gate with a photon-scattering error of $\ish 10^{-4}$, significantly below the error threshold for fault-tolerant operations~\cite{Fowler2012}. 

Similar experiments using trapped-ion qubits stored in two different elements (\Be{9} and \Mg{25}) have recently been carried out in the NIST Ion Storage Group~\cite{Tan2015}. Subsequent to the submission of our manuscript, a CHSH-Bell test which closes both detection and locality loopholes, using heralded entanglement of remote electron spins, has been reported~\cite{Hensen2015}.

\section*{METHODS}

{\noindent\bf Ion crystal order.} The \Ca{40}--\Ca{43} ion crystal ordering is kept constant during the experiments to control systematic errors. The principal error which would arise if the ion order were not controlled is due to an (undesired) axial magnetic field gradient that causes the magnetic field between the two ions to differ by $0.18\uT$. This means that the qubit frequencies for the two possible ion orders differ by $\approx 5\kHz$, which would lead to errors in single-qubit rotations. We measure the frequency of each qubit using slow (typically 100\us) carrier $\pi$-pulses, interleaved with the main experimental pulse sequence, which allows us to detect and to correct for both common-mode qubit frequency changes (due to drift in the global magnetic field $B$) and differential changes (due to incorrect ion crystal ordering). If the ion order is wrong, we randomly reorder the crystal until the order is correct with a short period of Doppler heating to melt the crystal, followed by a short period of Doppler cooling.
\\

{\noindent\bf Single-qubit phases and light shifts.} Despite the qubits having very different frequencies, no special phase control is needed to implement the entangling gate. The \Ca{43} qubit phase is tracked by the microwave local oscillator, and the \Ca{40} qubit phase is tracked by the difference phase of the co-propagating Raman beams, in turn referenced to a radio-frequency local oscillator. The phases of the Raman beams that implement the entangling gate have no relationship to either of the qubit phases. However, the travelling standing wave resulting from the interference of the Raman gate beams also generates an isotope-dependent differential light shift on each qubit with an amplitude that oscillates at the Raman difference frequency $\delta$. Over the course of the gate operation this light shift adds phase shifts to the qubits that depend on the (uncontrolled) optical phase difference of the Raman beams. These uncontrolled phase shifts reduce the fidelity of the gate operation. We greatly reduce this light shift error by shaping the turn-on and turn-off of the Raman laser intensities with a characteristic time of $1\us$; we estimate that without this pulse-shaping the light shift would lead to an average gate error of up to 5\% (see ref.\cite{Ballance2014a}).

We adjust the polarisation of each Raman beam individually to null the differential light shift from each single beam on the \Ca{40} qubit. (The interference of the two gate beams nevertheless gives rise to a polarization modulation which provides the state-dependent force.) Due to the difference in atomic structure there is a residual light shift on the \Ca{43} qubit of $\approx 0.2\%$ of the light shift for a purely circularly-polarised beam of the same intensity and frequency. This small light shift does not cause any significant issues in the experiments reported here; if necessary it could be suppressed further by increasing the Raman detuning at the expense of requiring more Raman beam power. 
\\

{\noindent\bf State preparation and measurement errors.} To perform individual single-shot qubit readout, we selectively shelve one qubit state of each ion to the \lev{3D}{5/2} level, then apply the Doppler cooling lasers sequentially in time first for one isotope, then for the other. If an ion was not shelved it fluoresces, and this is detected with a photomultiplier. We simultaneously shelve the two isotopes using a weak 393\nm\ beam resonant with the \Ca{43} $\hfslev{4S}{1/2}{4,+4}\leftrightarrow\hfslev{4P}{3/2}{5,+5}$ transition, with a 1.94\GHz\ EOM sideband which drives the \Ca{40} $\fslev{4S}{1/2}{+1/2}\leftrightarrow\fslev{4P}{3/2}{+3/2}$ transition. An intense 850\nm\ beam resonant with the \Ca{40} $\lev{3D}{3/2}\leftrightarrow\lev{4P}{3/2}$ transition makes the shelving for this isotope state-selective, through an EIT process involving the two transitions~\cite{McDonnell2004}. The \Ca{43} shelving is state-selective due to the $3.2\GHz$ splitting between the two qubit states~\cite{Myerson2008}. Both these shelving processes have a maximum theoretical efficiency of $\approx 90\%$ due to leakage to \lev{3D}{3/2} (which for \Ca{43} could be eliminated using a further 850\nm\ beam if required~\cite{Myerson2008}), leading to readout errors of $\bar{\epsilon}\approx 5\%$ when averaged over both qubit states. From independent experiments (similar to those we describe in ref.~\cite{Harty2014}), we estimate the state-preparation error to be $\approx 0.1\%$, which is negligible compared with the readout error. 

We measure the readout errors for each qubit state of each isotope, by preparing and measuring each state typically 10,000 times. Since the qubits are measured individually, it is then straightforward to calculate the linear mapping that corrects for the readout errors, provided that they remain constant. The readout errors relevant to the entangling gate experiment (figure~\ref{F:parity}) were measured to be $\bar{\epsilon}_{40}=7.7(2)\%$ for \Ca{40} and $\bar{\epsilon}_{43}=4.4(2)\%$ for \Ca{43} (averaged over both qubit states). Measurements of the readout errors were interleaved with the gate experimental runs, to check for systematic drifts, and were made using the mixed-isotope crystal, to avoid systematic effects associated with ion position. We estimate the systematic uncertainty in determining the readout errors to be $\approx 0.1\%$, less than the statistical error in these measurements. If we did not correct for readout errors, the apparent infidelity in the Bell state would increase by $\approx\frac{3}{2}(\bar{\epsilon}_{40}+\bar{\epsilon}_{43})\approx 18\%$. For the CHSH test, we do not correct for readout errors, but we nevertheless measure them in order to calculate the maximum attainable CHSH parameter ($S\sub{max}$). 

\newpage

{\noindent\bf Acknowledgements}
This work was supported by the U.K.\ EPSRC ``Networked Quantum Information Technology'' Hub and the U.S.\ Army Research Office (contract W911NF-14-1-0217). D.M.L.\ would like to thank Aida and Esther Andrade Castillo for their hospitality while revising the manuscript.
\\

{\noindent\bf Author contributions}
All authors contributed to the development of the apparatus and/or the design of the experiments. J.P.H.\ and D.M.L.\ conceived the experiments and took preliminary data. C.J.B.\ and V.M.S.\ designed and performed the experiments described here, analysed data and produced the figures. C.J.B.\ and D.M.L.\ wrote the manuscript, which all authors discussed.
\\

{\noindent\bf Author information}
Correspondence should be addressed to C.J.B.\ ({\tt c.ballance@physics.ox.ac.uk}) or D.M.L.\ ({\tt d.lucas@physics.ox.ac.uk}).

\end{document}